\documentclass[preprint,12pt]{elsarticle}

\usepackage{amsmath,amssymb,amsfonts}
\usepackage{bm}

\journal{Physica Medica}

\begin{document}

\begin{frontmatter}



\title{Generating spectral dental panoramic images from single energy computed tomography volumes}


\author{Villseveri Somerkivi\textsuperscript{a,b,c}, Anna Kolb\textsuperscript{a,b}, Thorsten Sellerer\textsuperscript{a,b}, Daniel Berthe\textsuperscript{a,b}, York H{\"a}misch\textsuperscript{d}, Tuomas Pantsar\textsuperscript{d}, Henrik Lohman\textsuperscript{d}, Franz Pfeiffer\textsuperscript{a,b,e,f}}

\affiliation{organization={Chair of Biomedical Physics, Department of Physics, School of Natural Sciences, Technical University of Munich},
            addressline={James-Franck-Stra{\ss}e 1}, 
            city={Garching},
            postcode={85748}, 
            state={Bavaria},
            country={Germany}}
            
\affiliation{organization={ Munich Institute of Biomedical Engineering, Technical University of Munich},
            addressline={Boltzmannstra{\ss}e 11}, 
            city={Garching},
            postcode={85748}, 
            state={Bavaria},
            country={Germany}}
            
\affiliation{organization={Planmeca Oy},
            addressline={Asentajankatu 6}, 
            city={Helsinki},
            postcode={00880},
            country={Germany}}

\affiliation{organization={Varex Imaging Corporation},
            addressline={1678 S. Pioneer Road}, 
            city={Salt Lake City},
            postcode={84104}, 
            state={Utah},
            country={USA}}
            
\affiliation{organization={Department of Diagnostic and Interventional Radiology, Klinikum rechts der Isar, Technical University of Munich},
            addressline={Ismaninger Straße 22}, 
            city={Munich},
            postcode={81677}, 
            state={Bavaria},
            country={Germany}}
            
\affiliation{organization={Institute for Advanced Study, Technical University of Munich},
            addressline={ Lichtenbergstra{\ss}e 2 a}, 
            city={Garching},
            postcode={85748}, 
            state={Bavaria},
            country={Germany}}

\begin{abstract}
\noindent
{\bf Purpose:} To implement a framework generating synthetic spectral panoramic images from single energy CT volumes. Using the framework output to compare the synthetic images against experimental spectral panoramic images for cross-verification. \\
{\bf Methods:} A simulation framework for generating synthetic spectral panoramic images from CT volumes is described. A cone beam CT scan of an anthropomorphic head phantom is used as input. An experimental spectral panoramic image of the same phantom is acquired. \\
{\bf Results:} The output of the framework of an anthropomorphic head phantom is compared against an experimental spectral panoramic image of the same phantom. The synthetic and experimental spectral panoramic images resemble each other considerably, especially the bone features. In the soft tissue images, there are some deviations, which are a result of the differences between the experimental and synthetic processing pipelines. \\
{\bf Conclusions:} It is demonstrated that generating synthetic spectral panoramic images from single energy CT volumes is possible. The synthetic images have many similarities with the experimental results, increasing the confidence in the correctness of the information contained within experimental spectral panoramic images and indicating that the synthetic images could be useful in further research. \\
\end{abstract}




\end{frontmatter}


\section{Introduction}

\begin{figure}[ht]
   \begin{center}
   \includegraphics[width=12cm]{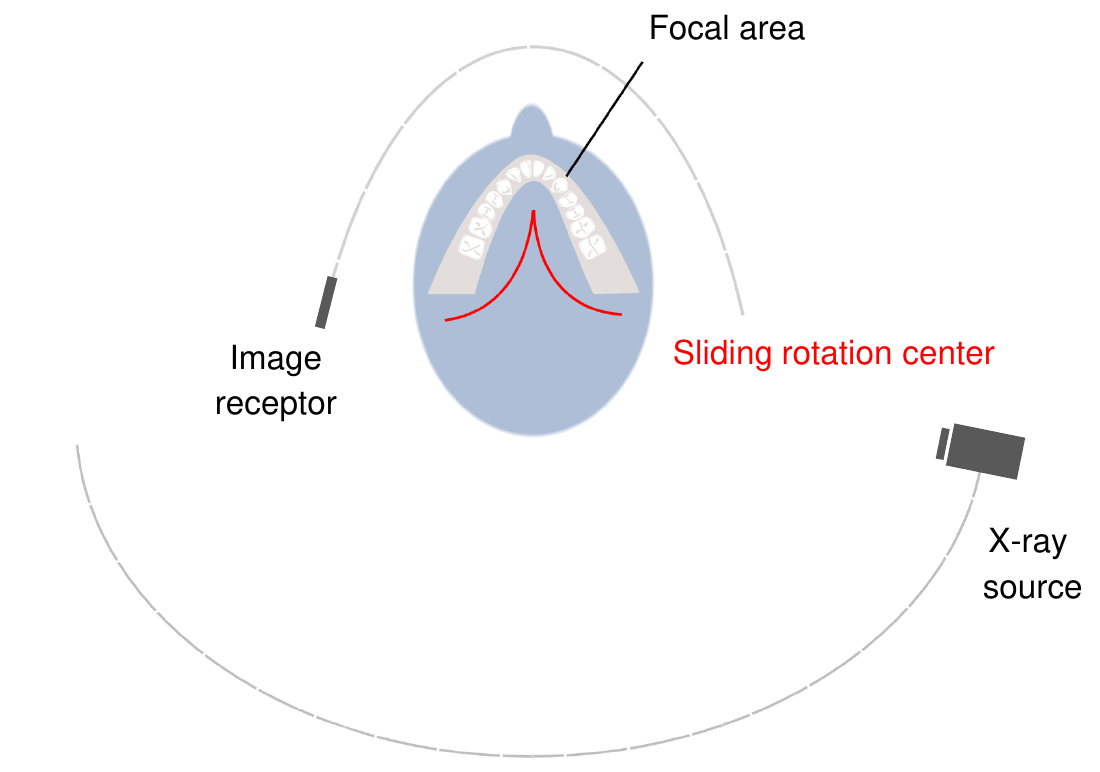}
   \caption{The working principle of panoramic imaging. Shown are the trajectories of the image receptor and the X-ray source. The position of these two decides the focal area (also called the sharp layer) and the rotation center, which in this case is sliding. The focal area aligns with the teeth. 
    }
   \label{fig_panScheme}
    \end{center}
\end{figure}

Dental panoramic X-ray imaging is a common imaging modality around the world. According to United Nations Scientific Committee on the
Effects of Atomic Radiation, the panoramic imaging examination frequency  per 1000 population was 140 in Germany, 120 in Japan, 74 in the United Kingdom, and 64 in the United States \cite{united22sources}. It provides a low-dose radiographic image, which is used mainly for a high- resolution view of the maxilla, the mandible, and the teeth and any lesions in these structures \cite{ianucci2012dental}. A panoramic image is generated by moving and rotating the X-ray tube and the detector in parallel to the dental arch while maintaining a constant magnification of the arch, as shown in fig. \ref{fig_panScheme}. In the case of a digital detector, frames are captured at a high frame rate and summed with a partial overlap to generate the panoramic image \cite{mcdavid1995digital}. Different tissue types are superimposed in conventional, transmission-based radiography, including panoramic imaging. This superimposition causes difficulties and confusion in the interpretation of panoramic images. For example, the oral cavity can be mistaken for a fracture in the mandible, even though the cavity is in the soft tissue \cite{perschbacher2012interpretation}.

Spectral X-ray imaging has been studied extensively in the past 15 years \cite{schlomka2008experimental} \cite{alvarez2011estimator} \cite{ehn2017basis} \cite{flohr2020photon}. These studies have demonstrated that material decomposition provides additional information on top of the normal attenuation images by revealing clinical features which would be invisible in a conventional attenuation image. Recently, spectral panoramic imaging specifically has also been investigated \cite{fukui2015usefulness} \cite{langlais2015cadmium} \cite{somerkivi2023spectral}. A previous work demonstrated that it is possible to achieve material decomposition in panoramic imaging for as long as there is a sufficient number of counts for calibration and decomposition \cite{somerkivi2023spectral}. 

Currently, no clinical studies on the feasibility of spectral panoramic imaging have been published. Preliminary results from a previous pre-clinical study have been obtained \cite{somerkivi2023spectral}. However, the confidence in the results would increase if there was a reference to compare these results to. Several approaches to generate synthetic panoramic images from CT have been described \cite{bae2019semi} \cite{luo2016automatic} \cite{papakosta2017automatic} \cite{yun2019automatic}. These approaches aim at automatically or semi-automatically finding the dental arch, e.g. by using the maximum intensity projection along the axial slices followed by subsequent thresholding and morphological thinning. After the dental arch has been extracted, the panoramic image is generated by extracting voxel values along lines which are orthogonal to the dental arch. The length of lines is set so that the lines run through the mandible, the maxilla, and the teeth while the rest of the lower head is left out. This kind of process may be suitable for generating synthetic panoramic images for reading in dental practice. However, if verification is desired, then the whole image generation process should be modeled. This includes forward projecting actual frames and summing the frames to get the panoramic image. Additionally, the generation of basis material images is required when modeling spectral imaging.

This work introduces a framework for generating spectral panoramic images from CT volumes. We use a CT image of an anthropomorphic head phantom as input and generate a set of spectral panoramic images using the framework. We then cross-compare the output of the framework against experimental spectral panoramic images.

\section{Methods}

\subsection{Simulation framework}

\begin{figure}[ht]
   \begin{center}
   \includegraphics[width=12cm]{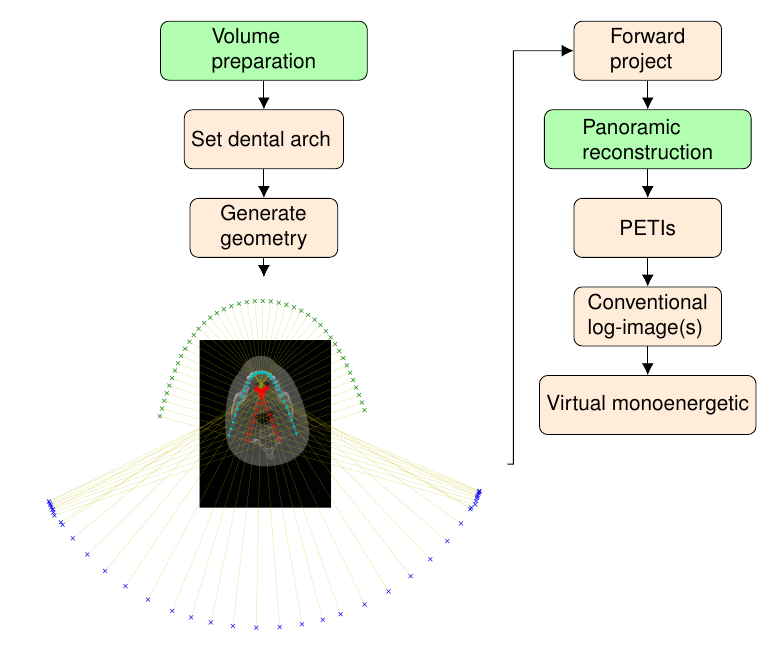}
   \caption{Main steps of the simulation framework. PETI stands for panoramic equivalent thickness image. The dental arch used and the geometry generated from the arch are shown overlaid with an axial slice of a volume on the bottom left side. Every 60th projection is shown, with green being the detector, dark blue the source, cyan the focal through, and red the rotation center.
    }
    \end{center}
   \label{fig_workflow}
\end{figure}

The framework was implemented in Python (Python Software Foundation, Wilmington, Delaware, United States). A proprietary X-AID module (MITOS GmbH, Munich, Germany) was used for the forward projection step. The main steps of the workflow are shown in fig. \ref{fig_workflow} and the detailed explanation is below.

The first part of the framework deals with preparing the input CT volume for the workflow. If the voxels are not isotropic, e.g. the input is from a helical CT, an interpolation is first carried out so that the output is isotropic. Secondly, the user has to rotate the volume along the three axes so that the volume is symmetric along the coronal and axial views, and the dental arch is horizontal in the sagittal view. The final step of the preparation is thresholding the volume to two basis materials. There are two ways to carry out the thresholding, depending on the type of the input volume.

A simple threshold is used in the segmentation if the imaging system's Hounsfield Unit (HU) values are not perfectly reliable due to scatter contamination \cite{sisniega2013monte}. A threshold of 1280 HU was used in this study. Both basis material images are scaled with their respective mean signals so that the mean value of the segmented volumes is 1.

If the HU values of the input volume are more reliable e.g., by reducing the scatter contamination by using an anti-scatter grid \cite{altunbas2021unified}, then a more sophisticated thresholding is possible in the framework. The method uses a sheetness score combined with a threshold in order to better segment the bone, as bones in the skull mainly consist of sheet-like structures \cite{cuadros2019mandible}. The sheetness score $S$ is calculated as in \cite{cuadros2019mandible}:

\begin{equation}
S = \text{ exp}\left[\frac{-R_\text{sheet}^2}{2\alpha^2}\right] \left(1-\text{exp}\left[\frac{-R_\text{tube}^2}{2\beta^2}\right]\right) \left(1-\text{exp}\left[\frac{-R_\text{noise}^2}{2\gamma^2 }\right]\right)
\label{eq_sheetness}
,
\end{equation}
where $R_\text{sheet}=\vert \lambda_2 \vert / \vert \lambda_3 \vert$, $R_\text{tube}=\vert (2\vert \lambda_3 \vert - \vert \lambda_2 \vert) - \vert \lambda_1 \vert ) \vert/ \vert \lambda_3 \vert$, and $R_\text{noise}=\sqrt{\lambda_1^2 + \lambda_2^2 + \lambda_3^2}$ with $\lambda_i$ being the local Hessian eigenvalues so that $\vert \lambda_1 \vert \leq \vert \lambda_2 \vert \leq \vert \lambda_3 \vert$. $\alpha$ and $\beta$ are set to 0.5 here, while $\gamma$ is equal to maximum of $R_\text{noise}$.  The output HU values $\Omega$ are then:

\begin{equation}
\Omega = \begin{cases}
-1000 &\text{if } (\Omega_{i}S\cdot5 +\Omega_{i}) \leq T \\
\Omega_{i}S &\text{othwerwise} \\
\end{cases}
,
\label{eq_th}
\end{equation}


where $\Omega_{i}$ are the HU values of the input volume and $T$ is the threshold value. $(\Omega_{i}S\cdot5 +\Omega_{i})$ was heuristically found to yield good results in the thresholding.

The next step is to trace the dental arch of the volume. The dental arch needs to be correctly registered in order for the panoramic reconstruction to succeed (c.f. fig \ref{fig_panScheme}). This requires the user to manually provide the coordinates of the dental arch. There are two ways to achieve this in the framework: u-shaped (9 coordinates) and exact (11 coordinates) arches. The u-shaped arch generates a smoother profile with a smaller number of rotations. The u-shaped arch was used in this work (shown in fig. \ref{fig_workflow}) as it corresponds better to a realistic trajectory. It is also possible to restrict the overlap of the mandible from the other side in the framework. This feature was not used for the results presented in this paper, as it was desirable to have the simulation results match the experimental results. The dental arch is created by interpolating between the user-given points by using the piecewise cubic hermite interpolating polynomial (PCHIP) algorithm. The algorithm is desirable for creating the dental arch, as it is more resistant to ringing artifacts than spline or regular polynomial-based interpolation algorithms \cite{emmendorfer2020conservative}.

After the dental arch is complete, the subsequent step is forward projecting the frames. The X-AID module uses a ray-driven forward projector which traces the path from the source to the target pixel and sums up everything along the path. The panoramic reconstruction is carried out by summing the frames according to the movement profile generated using the dental arch (c.f. fig. \ref{fig_workflow}).

The output of the panoramic reconstruction is called a panoramic equivalent thickness image (PETI) in this framework. In this case, the volume was thresholded into soft tissue and dentin. Thus, the output here is two PETIs of soft tissue and dentin with a unit of thickness (mm by default).

For the generation of log images, dentin and soft tissue PETIs were used as basis materials. The linear attenuation coefficients (LACs) are obtained from the XCOM database \cite{hubbell1995tables}. The pyXSFW framework \cite{malecki2012quantitative} was used to model the tungsten anode spectrum. The detector was modeled as having a sensor thickness of 0.75 mm of cadmium telluride, with each absorbed photon inducing one count. With the spectrum and the LAC(s) known for both basis materials and the detector, the output signal can be calculated:

\begin{equation}
I = \sum_{E=1}^{\infty}I_0(E)D(E)F(E)e^{-A_{st}\mu_{st}(E)-A_{d}\mu_{d}(E)},
\label{eq_bl}
\end{equation}

where $I_0(E)$ is the input spectrum, $D(E)$ is the detector absorption efficiency, $F(E)$ is the spectral response of the detector, $A_{st}$ and $A_{d}$ are the respective soft tissue and dentin line integral lengths, while $\mu_{st}(E)$ and $\mu_{d}(E)$ are the energy-dependent soft tissue and dentin LACs, respectively. 

No quantum noise was added to the images in this work, as the purpose was to evaluate the feasibility of the results in the sense that the output is as expected. While the experimental results are subject to quantum noise, it would not make the comparison easier or better to add noise to the synthetic images. If there was a need to add quantum noise, it should be added to the forward projected frames before panoramic reconstruction. This would ensure that the reconstructed image has a realistic noise distribution.

Virtual monoenergetic images (VMIs) are generated by multiplying the PETIs with their respective LACs and then summing them together.

The individual projections making up a single output pixel in a panoramic image all pass through the patient at a slightly different path. However, all overlap in the sharp layer (c.f. fig. \ref{fig_panScheme}). Anatomical noise in a panoramic image pixel is measured by recording all of the pixel values from the frames contributing to the pixel in the panoramic image. In other words, this indicates how different the individual line integral paths making up the final panoramic image pixel are. The mean of these values is the final PETI output value in the pixel.

\subsection{Experimental measurements}

\subsubsection{CBCT acquisition}

Planmeca Viso G7 (Planmeca Oy, Helsinki, Finland) was used for the cone beam CT (CBCT) acquisition. The acquisition was carried out in offset mode, and two rotations were conducted and stitched together to acquire the whole head phantom. The reconstructed volume was used as input to the simulation framework. The size of the reconstruction output grid in voxels was 1216x1193x1361 (width, lenght, height). The acquisition parameters are listed in table \ref{tab1}. The dose is higher than what would be used in the typical patient acquisition, as high resolution with lower noise was desired.

\begin{table}[htbp]
\begin{center}
\caption{CBCT acquisition parameters.}
\label{tab1}
\begin{tabular} {|c|c|}
\hline
source to image distance& 
70$\,$cm \\
acceleration voltage&
120$\,$kVp \\
tube current& 
11$\,$mA \\
exposure time& 
11.4$\,$s \\
dose-area product& 
4100$\,$mGycm$^2$ \\
magnification& 
1.5 \\
isotropic voxel size&
225$\,$$\mu$m \\
field of view (width$\times$height)&
25$\times$30$\,$cm$^2$ \\
\hline
\end{tabular}
\end{center}
\end{table}

The anthropomorphic head phantom RANDO SK150 (Radiation Analogue Dosimetry System; The Phantom Laboratory, Salem, NY, USA) was used in this study. The phantom is a skull encapsulated in plastic. The plastic has air cavities that mimic the upper respiratory tract. A hole has been drilled into the phantom from the bottom of the jaw so that there is also a cavity in the bottom of the jaw in the plastic and in the mandible.

\subsubsection{Spectral panoramic acquisition}

For the experimental data, a low-order polynomial model of line integrals for polychromatic X-ray spectrum is used \cite{cardinal1990accurate}. The expected number of counts $\hat{y}_i^s$ is then

\begin{equation}
\label{fwd_model_decomp}
\hat{y}_i^s = b_i^s e^{- \hat{l}_i^s}, \ \hat{l}_i^s = P_i^s(A_i; c_{is}), \ A_i = {(A_i^0,...,A_i^n)}^T,
\end{equation}
where $b_i^s$ is the flatfield signal for pixel $i$, spectral measurement $s$ and $A_i$ are the basis material line integrals. For this paper, the polynomial model $P_i^s$ was chosen to have two basis materials:
\begin{equation}
\begin{split}
P_i^s & = \frac{N}{D}  \\
N & = c_{0} + c_1 A_i^0 + c_2 A_i^1 + c_3 (A_i^0)^2 + c_4 (A_i^1)^2 +c_5 A_i^0 A_i^1  \\
D & = 1 + c_6 A_i^0 + c_7 A_i^1
\label{2mat_poly}
\end{split}
,
\end{equation}
where $c$ are the fit coefficients. Here, the subscript is the number of the fit coefficient, while pixel and spectral measurement subscripts are omitted.

Calibration measurements are least-squares fitted to establish the fit parameters $c_{is}$:
\begin{equation}
\label{cal_poly}
c_{is} = \text{arg min} \sum_{k=1}^K w_k  {\left(  l_{ik}^s - P \left( A_{ik}; c_{is} \right) \right)}^2,
\end{equation}
where $w_k$ are the weights with $k$ being the index of the calibration measurement, $l_{ik}^s$ the linearized calibration measurements,  and $A_{ik}$ the line integrals. The standard deviation of the particular calibration measurement determines the weights.

After the forward model has been calibrated, a maximum likelihood estimation is carried out to find
the basis materials thicknesses $A_i$, which maximize the likelihood given the spectral measurements. Assuming independent Poisson statistics for the spectral measurements, minimizing the corresponding negative log-likelihood is  computationally more efficient:
\begin{equation}
\label{ml_decomp}
\mathcal{A}_i = \text{arg min}  - \mathcal{L}_i(A_i) = \sum_{s=1}^S \hat{y}_i^s(A_i) - y_i^s \text{ln}(\hat{y}_i^s(A_i) ). 
\end{equation}

A dictionary learning-based regularization is also applied to the decomposition to reduce noise, with details provided in \cite{mechlem2018spectral}.

Panoramic imaging presents a new challenge to the decomposition process, as individual frames in panoramic imaging have a very low number of counts. Directly decomposing frames would induce severe statistical bias to the decomposition \cite{mechlem2017joint}. On the other hand, the count statistics in the final panoramic image are sufficient for typical line integrals in the jaw area \cite{somerkivi2023spectral}. However, to decompose the final panoramic image instead of individual frames, an appropriate calibration needs to be carried out. One such way is generating virtual panoramic images for calibration. This involves taking still calibration exposures, taking the mean of each measurement, and performing a panoramic reconstruction for each mean frame by using the mean frame as every frame in the reconstruction. The the final panoramic image can then be calibrated using these virtual panoramic images. Thus, the actual panoramic exposure will have the same flatfield values and pixel weights as the calibrated panoramic image. A change of basis from the calibration materials into soft tissue and dentin is carried out after the decomposition. A detailed explanation of the whole process can be found in \cite{somerkivi2023spectral}. 

An experimental spectral photon counting panoramic system was used to capture an image of the same phantom. The Planmeca Promax Mid (Planmeca Oy, Helsinki, Finland) system has a DC-Vela detector (Varex Imaging Corporation, Salt Lake City, USA). DC-Vela is a photon counting detector with two adjustable thresholds, a charge sharing correction, and a 0.75 mm cadmium telluride sensor layer. The acquisition parameters for the experimental panoramic system are shown in table \ref{tab2}.

\begin{table}[htbp]
\begin{center}
\caption{Panoramic acquisition parameters.}
\label{tab2}
\begin{tabular} {|c|c|}
\hline
source to image distance& 
60$\,$cm \\
acceleration voltage&
70$\,$kVp \\
tube current& 
9$\,$mA \\
exposure time& 
18$\,$s \\
dose-area product& 
72.1$\,$mGycm$^2$ \\
magnification& 
1.4 \\
field of view (width$\times$height)&
0.6$\times$15.41$\,$cm$^2$ \\
\hline
\end{tabular}
\end{center}
\end{table}

A decomposition process like the one described in this section was not modeled in the simulation framework. If the same forward model is used to simulate the transmission and to decompose, the output of the decomposition will be the same as the input in the absence of noise. This holds as long as the input volume is segmented into exactly two basis materials.
	     
\section{Results}

\begin{figure}[h!]
   \begin{center}
   \includegraphics[width=12cm]{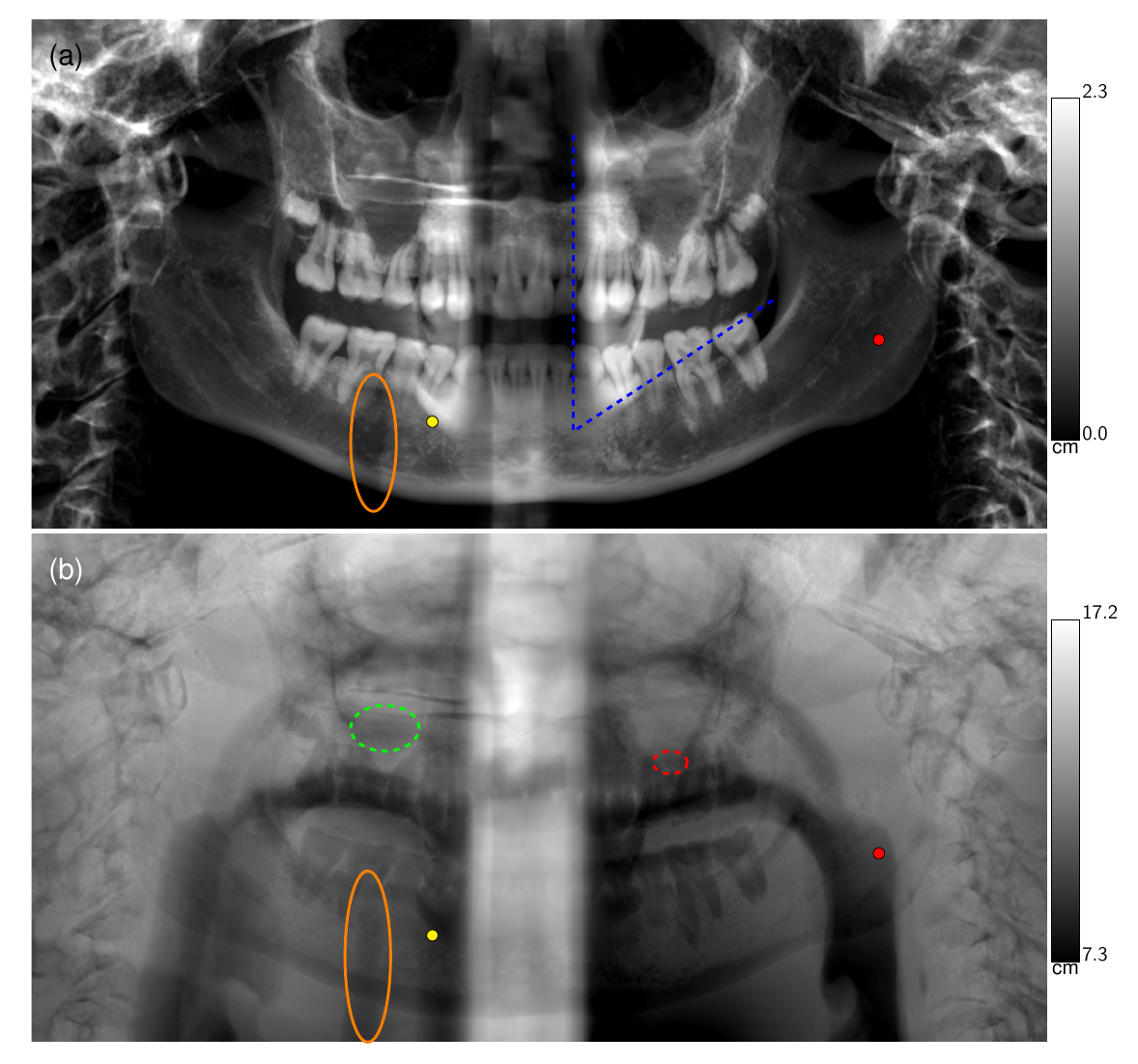}
   \caption{Synthetic pan images, generated from the CBCT scan of the RANDO SK150 phantom. A solid orange ellipse highlights the drilled hole in both images. The blue dashed lines highlight the boundary for the overlapping mandible from the other side in the bone image (a). The two dashed ellipses show small air gaps in the plastic in the soft tissue image (b). Many dark features in the soft tissue image (b) are not air but the absence of bone i.e., there is a strong negative correlation between the two images.}
   \label{fig_panSim}
    \end{center}
\end{figure}

The output of the simulation framework is shown in fig. \ref{fig_panSim}. As expected, the maxilla, the mandible, and the teeth are the prominent features of the dentin image (a). The spine is also visible to the sides. The soft tissue image (b) has larger line integral thicknesses to the center. Outside of the center, there are homogeneous areas with, darker air gaps with the upper respiratory tract being the most prominent. The tract starts at the sides, next to the spine, and extends upwards until splitting into the oral and nasal tracts and arching towards the center. There are also darker areas in the soft tissue image (b) in the same places where the bone features are in the dentin image (a). For example, the spine, the mandible, and the teeth are visible as darker areas in the soft tissue image (b). The drill hole (solid orange ellipse) is visible in both images. This is expected, as the hole was indeed drilled through the bottom of the jaw and partially through the mandible. The air gaps highlighted in (b) by the larger and smaller dashed ellipses are not part of the respiratory tracts. The gaps are roughly where the maxillary sinuses would be. While darker areas are inside the ellipses, the areas are only barely visible.

\begin{figure}[h!]
   \begin{center}
   \includegraphics[width=12cm]{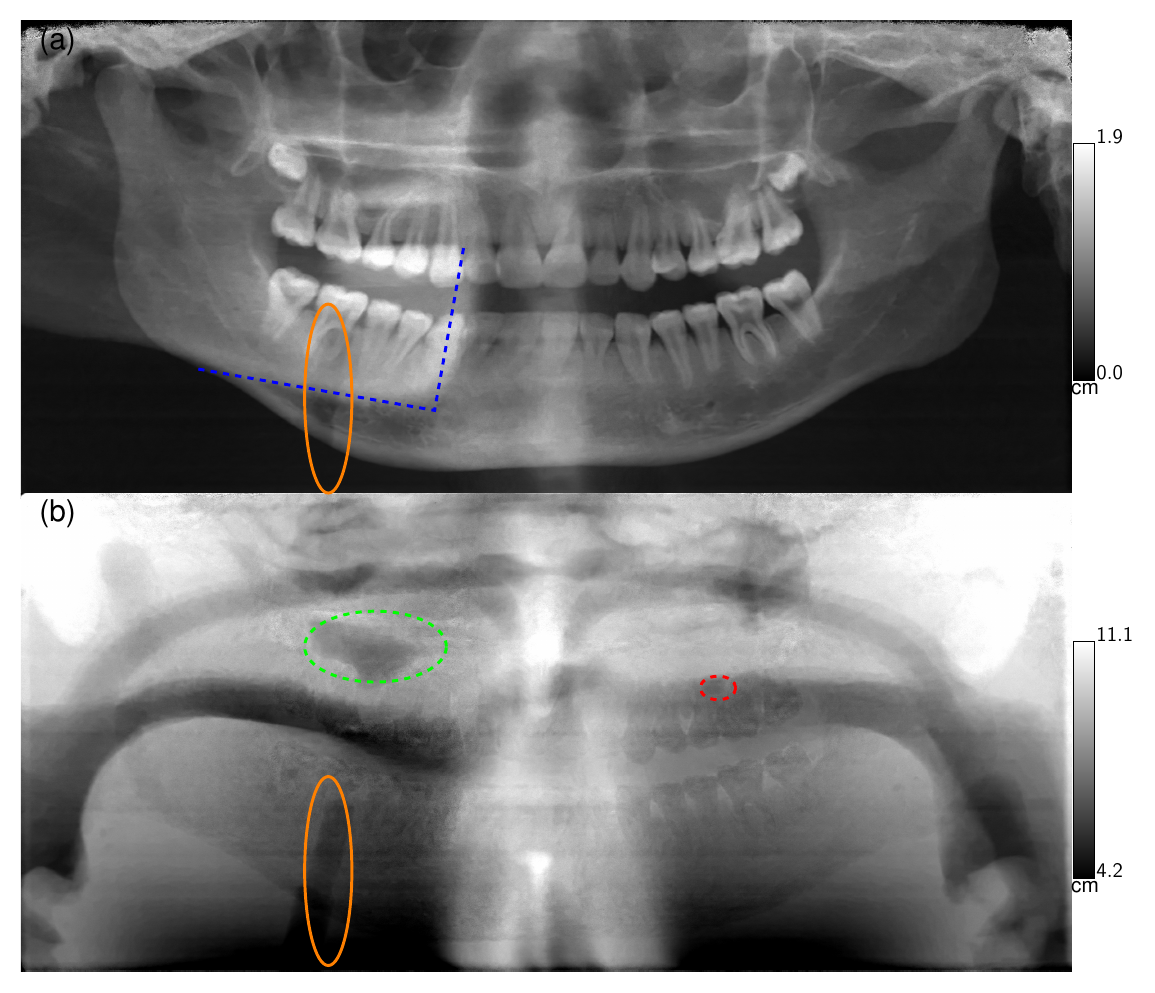}
   \caption{The experimental panoramic images. The highlighted details are the same as in fig. \ref{fig_panSim}. The overlap from the mandibles is not symmetric in (a), as the overlap on the right side of the image only starts from a higher and more away from the center than the overlap on the left side. Here, the darker features in (b) are mostly air gaps in the phantom. The yellow and the red points are the anatomical noise measurement pixels.}
   \label{fig_panExp}
    \end{center}
\end{figure}

The basis material images from the experimental system are shown in fig. \ref{fig_panExp}. The maxilla, the mandible, and the teeth are the dominant features in the dentin image (a). The soft tissue image (b) is quite homogeneous, with the air gaps, especially the upper respiratory tract being the dominant features. The air gaps not being part of the respiratory tract are again highlighted by the dashed ellipses. In the experimental image \ref{fig_panExp}, these gaps stand out more prominently, although the smaller gap overlapping with the lower nasal cavity limits its visibility.

Figure \ref{fig_panExp} is similar to what was described in a previous work \cite{somerkivi2023spectral}. The most notable difference is the presence of the upper respiratory tract in the soft tissue image. Note that the images of this work are left-right flipped, unlike the image in the previous work. Dentin was considered to be a good basis material in the previous work, so it was kept the same in this work. The calibration of the panoramic system, including the results of the test decomposition and noise considerations are discussed in detail in a previous work \cite{somerkivi2023spectral}.

There are some differences in features between the bone component in figures \ref{fig_panSim} and \ref{fig_panExp}. The mean line integral length of fig.3 (a) is lower than fig. 4 (a). The overlap from the mandible from the other side has locally very large values in fig. \ref{fig_panSim}, however. This is why the windowing is wider in fig. \ref{fig_panSim} (a) than in fig. \ref{fig_panExp} (a). In fig. \ref{fig_panSim} (a), the contrast between the body and the base of the mandible is more prominent than in \ref{fig_panExp} (a).

The soft tissue images (b) have notable differences in \ref{fig_panSim} and \ref{fig_panExp}. The soft tissue image in \ref{fig_panExp} is much more homogeneous than \ref{fig_panSim}. More specifically, the bright features in the synthetic dentin image (a) in fig. \ref{fig_panSim} appear dark in the soft tissue image (b) in fig. \ref{fig_panSim}.  In other words, the negative correlation between the experimental basis material images in \ref{fig_panExp} is much smaller than in the synthetic images in fig. \ref{fig_panSim}. On the other hand, the upper respiratory tract is clearly visible in both soft tissue images. Additionally, the drill hole can be seen in both the dentin images and both soft tissue images in figures \ref{fig_panSim} and \ref{fig_panExp}.

The shape of one of the overlapping mandibles is highlighted by a blue dashed line in figures \ref{fig_panSim} and \ref{fig_panExp}. But the exact location and shape of the overlap is different between the two figures. This is a result of different imaging geometries. The experimental device uses a fixed geometry, whereas the simulator uses a manually defined geometry, which is specifically tuned for each volume.

\begin{figure}[h!]
   \begin{center}
   \includegraphics[width=12cm]{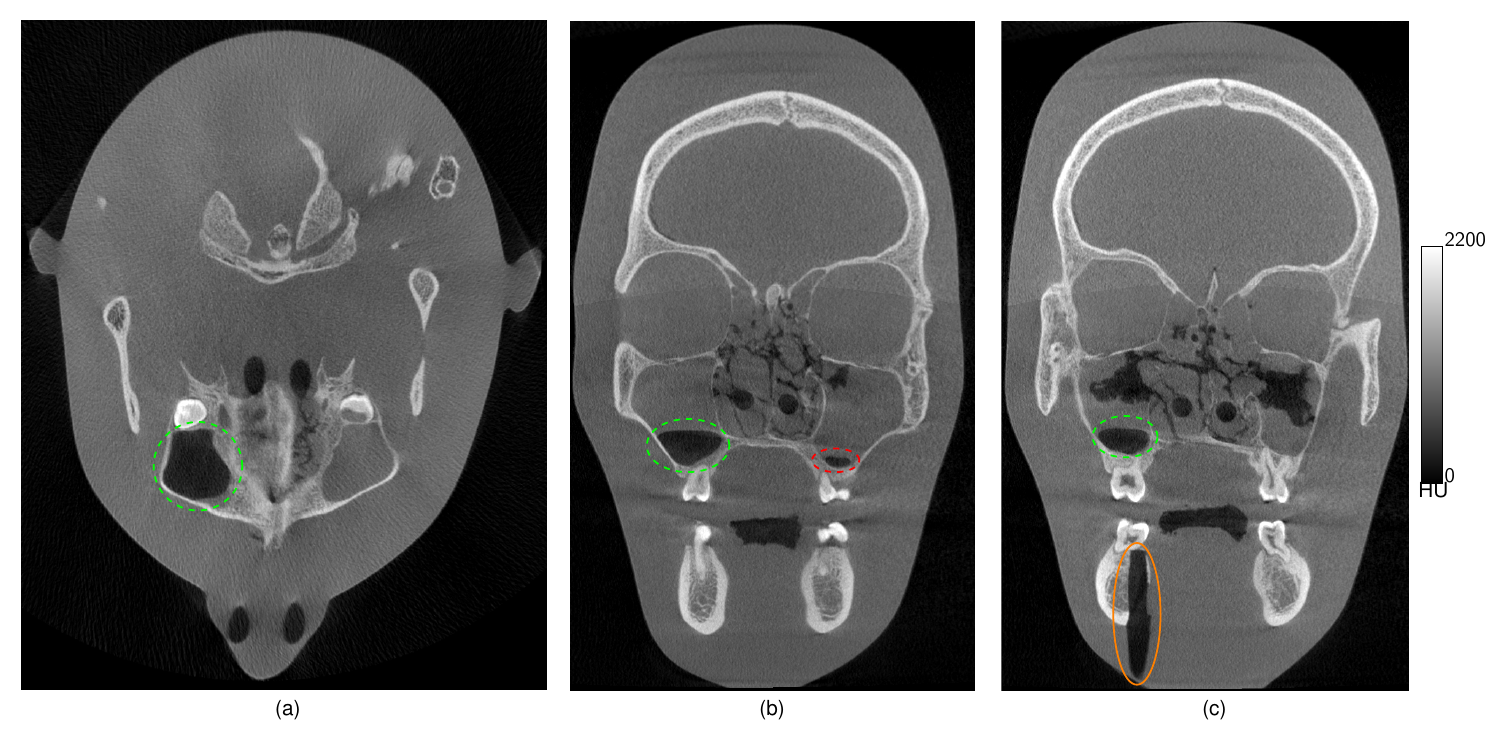}
   \caption{An axial and two coronal slices showing the same details in the CBCT image as in the two pan images. From the CBCT image, it is confirmed that there are indeed air gaps in the highlighted positions. The visible soft tissue boundary in (b) and (c) around eye sockets' upper part marks the two rotations' stitching boundary.}
   \label{fig_cbct} 
    \end{center}
\end{figure}

The details in fig. \ref{fig_cbct} show slices from the CBCT acquisition. The slices confirm that there are indeed air gaps in the positions highlighted in fig. \ref{fig_panSim} and fig. \ref{fig_panExp}. The drill hole is inside the mandible but is visible in all of the images in figures \ref{fig_panSim} and \ref{fig_panExp}.

\begin{figure}[h!]
   \begin{center}
   \includegraphics[width=12cm]{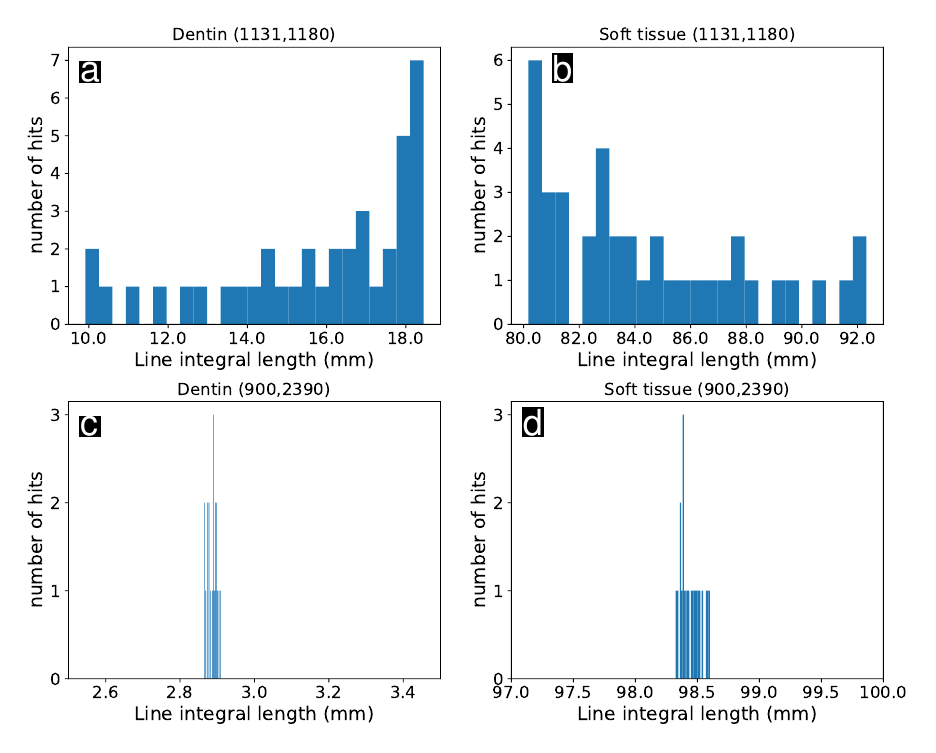}
   \caption{dentin and soft tissue anatomical noise histograms from two locations with pixel indices given in (height, width). The yellow point in fig. \ref{fig_panSim} corresponds to the top row (a) and (b)  and the red point (c) and (d) to the bottom row. Observe that the x-scales are very different for the two rows (compare (a) vs (c) and (b) vs (d)).}
   \label{fig_hists}
    \end{center}
\end{figure}

Figure \ref{fig_hists} displays the anatomical noise for both the soft tissue and the dentin PETIs of two pixels in fig. \ref{fig_panSim}. More precisely, each histogram in fig. \ref{fig_hists} is generated by recording the individual projections contributing to the panoramic image's particular pixel. These two cases were included to highlight the extremes in the anatomical noise. The upper two histograms are near the boundary of the mandible overlap. The values close to 18 mm (b) are a sum of the overlap and the actual signal, while the lower values are only from the actual signal. The lower two histograms are from a region with little anatomical noise. Thus, the values are very close to each other. Most of the panoramic image has a noise magnitude close to pixel (900, 2390), while an increase in the noise is observed for bone -- soft tissue boundaries with pixel (1131, 1180) as an extreme case with the mandible overlap.

\section{Discussion}

The presented framework enables generation of panoramic images from CT and CBCT volumes. The PETIs generated this way resemble the experimental material decomposition panoramic images. The framework also enables quantitative and pixel-wise estimation of anatomical noise in panoramic images.

The thresholding approach for segmentation assumes that one voxel induces a signal only on the soft tissue or bone image, but not both. The assumption may be correct for voxels containing only plastic, as they should only contribute to the soft tissue image. The skull, on the other hand, consists of more than one type of tissue. For instance, cancellous bone \cite{hasgall2018database}, cortical bone \cite{hasgall2018database}, dentin \cite{zenobio2011elemental}, and enamel \cite{zenobio2011elemental} have different effective atomic numbers and are to an extent present in the skull. Furthermore, in reality, one voxel can have more than one tissue type. 

The overlap of the mandible from the other side appears different in figures \ref{fig_panSim} and \ref{fig_panExp} due to the different geometries. The difference has an impact on both location and the intensity of the overlap. Ideally, the geometry of the experimental setup should also be used in the simulation framework to mitigate these differences. However, the exact geometry of the used imaging system is not publicly available. The non-symmetrical overlap of the mandibles in the experimental image is likely caused by a sub-optimal mounting of the detector in the experimental system. 

The values of a single panoramic image output pixel consist of line integrals passing through the lower head at slightly different paths. The values of the line integrals passing through the mandible are very close to each other (bottom row of fig. \ref{fig_hists}) (excluding the opposing mandible overlap region as seen on top row of fig. \ref{fig_hists}). This implies that a minor deviation in the angle or location does not change the line-integral value notably. Thus, a slight variation in the geometry should not provide notably different values assuming the boundary condition that the sharp layers of the two geometries mostly overlap. This condition is satisfied in this work as details like the pulps or the mandibular canals would not be visible anymore if the the sharp layers of the two geometries were not close to each other.

The choice of PETI materials is somewhat arbitrary when segmenting CT volumes. Dentin and soft tissue were used in this study as they are a good guess with respect to the effective atomic numbers and densities of the underlying materials. Cortical bone could have been used instead of dentin, as their attenuation profiles are close \cite{hasgall2018database, zenobio2011elemental}. The experimental setup decomposes images initially into the calibration materials, in this case, titanium and polyoxymethylene (POM). In the subsequent change of basis it has to be decided which materials and energies for low and high bins are used, which are again arbitrary to some extent. In other words, there is no single correct value for the experimental output images. Thus, it is not feasible to expect the synthetic and experimental panoramic images to match exactly. 

The input volumes should ideally have both high-resolution and reliable HU values. This would enable reliable two-material segmentation while preserving bone feature visibility. Reliable HU-values would also increase the volume of segmented images, as with the system used in this study some of the cancellous bone was segmented as soft tissue. Partially missing cancellous bone is a plausible explanation as to why the contrast between the body and the base of the mandible is stronger in fig. \ref{fig_panSim} than in fig. \ref{fig_panExp}. An increase in the soft tissue -- bone threshold would result in soft tissue being segmented as bone. With reliable HU values, the bone could be potentially segmented into more than two basis materials, making the results even more realistic.

With the aforementioned differences in the generation of the experimental and synthetic spectral panoramic images in mind, it was expected that the two images are not exactly the same. Nevertheless, the close resemblance of the two images increases the confidence that the output of the experimental spectral panoramic imaging workflow is feasible and expected. Additionally, the developed simulation framework will be useful in further exploring spectral panoramic imaging.

The framework will be adapted for dual-energy CT (DECT) input in a future work. DECT images can be decomposed and forward projected directly without thresholding. This way, a voxel can consist of more than one material. DECT decomposed images can also be used as a reference for estimating bone mineral density (BMD) \cite{laugerette2019dxa}. Photon counting CT (PCCT) provides the same benefits as DECT while having spatial resolution close to CBCT \cite{thomsen2022effective}. Thus, PCCT would be the ideal way to provide the ground truth for BMD estimation from spectral panoramic images. Another future work will explore the clinical relevance of PETIs.

\section{Conclusion}

A framework for generating synthetic spectral panoramic images from CT volumes has been successfully implemented. The synthetic panoramic images obtained using the framework support the results obtained from the experimental spectral panoramic system. The framework also provides a basis for further research.

\section*{Acknowledgements}
The authors acknowledge financial support through the EQAP project, which is part of the One-Munich Strategy. The authors acknowledge Varex Imaging Corporation for the hardware support with the photon counting detector and Planmeca Oy for the support with the experimental setups. The authors acknowledge Ari Hietanen and Juha Koivisto for helpful discussions.

\section*{Conflicts of interest}
VS is an employee of Planmeca Oy. YH, TP, and HL are employees of Varex Imaging Corp. The remaining authors declare no conflicts of interest.



\bibliographystyle{elsarticle-num} 





\end{document}